\documentclass[10pt]{article} 

\usepackage[english]{babel}
\usepackage[utf8]{inputenc}

\usepackage{float}		
\usepackage{times}		
\usepackage{graphicx}		
\usepackage{url}
\usepackage{lscape}
\usepackage{ulem}
\usepackage{epstopdf}
\usepackage{longtable}
\usepackage{amsmath}
\usepackage{amssymb}
\usepackage{multicol}
\usepackage[font={small}]{caption}
\usepackage{subcaption}
\usepackage{abstract}
\usepackage{fancyvrb}
\usepackage{verbatim}
\usepackage[procnames]{listings}
\usepackage[bottom]{footmisc}
\usepackage{tabto}
\usepackage{paralist} 
\usepackage{xcolor}
\usepackage[document]{ragged2e}
\usepackage{txfonts} 
\usepackage{microtype}
\usepackage{enumitem}  
\usepackage{comment} 
\usepackage[T1]{fontenc}
\usepackage{wrapfig}
\usepackage{tabularx}

\usepackage{placeins}
\usepackage{caption}
\usepackage[numbers]{natbib}
\usepackage{bm}
\usepackage{listings}
\usepackage{hyperref}

\hypersetup{
    colorlinks,
    linkcolor={blue!70!black},
    citecolor={blue!70!black},
    urlcolor={blue!70!black}
}

\usepackage{mathtools}
\usepackage{physics}
\usepackage{nicefrac}
\usepackage{latexsym,amsfonts}
\numberwithin{equation}{section}

\newcommand{\sgra}{Sgr\,A$^*$ }

\hypersetup{
    colorlinks,
    linkcolor={blue!70!black},
    citecolor={blue!70!black},
    urlcolor={blue!70!black}
}

\usepackage{imakeidx}

\makeindex

\usepackage[nottoc]{tocbibind}

\DeclareGraphicsExtensions{.pdf,.png,.jpeg}

\def\memonr{2025-TDWG-01}

\title{\vspace{-1.25cm} 
\textcolor{gray}{\textit{EHT Time Domain Working Group}}\\
\vspace{0.40cm}
\textbf{A Novel Calibration and Imaging Method for ALMA Observations of \sgra}
}

\author{  
Ezequiel Albentosa-Ruiz$^{1}$,
Iv\'an Mart\'i-Vidal$^{1,2}$,
Ciriaco Goddi$^{3,4,5,6}$,
Alejandro Mus$^{5}$
}

\def\affiliations{
\hspace{0.5cm}$^{1}$\textit{Dpt. Astronomia i Astrof\'isica, Universitat de Val\`encia, C/ Dr. Moliner 50, 46120 Burjassot, Spain} \\
\hspace{0.5cm}$^{2}$\textit{Observatori Astron\`omic, Universitat de Val\`encia, C/ Cat. Jos\'e Beltr\'an 2, 46980 Paterna, Spain} \\
\hspace{0.5cm}$^{3}$\textit{Universidade de S\~ao Paulo, Instituto de Astronomia, Geof\'isica e Ci\^encias Atmosf\'ericas, Departamento de Astronomia, S\~ao Paulo, SP 05508-090, Brazil} \\
\hspace{0.5cm}$^{4}$\textit{Dipartimento di Fisica, Universit\'a degli Studi di Cagliari, SP Monserrato-Sestu km 0.7, I-09042 Monserrato, Italy} \\
\hspace{0.5cm}$^{5}$\textit{INAF - Osservatorio Astronomico di Cagliari, via della Scienza 5, I-09047 Selargius (CA), Italy} \\
\hspace{0.5cm}$^{6}$\textit{INFN, Sezione di Cagliari, Cittadella Univ., I-09042 Monserrato (CA), Italy} \\
}

\date{March 07, 2025 -- Version 1.1}

\begin{document}

\begin{figure}[!t]
\begin{minipage}[t]{0.49\textwidth}
    \vspace{-0.13\linewidth}{\includegraphics[width=0.6\linewidth]{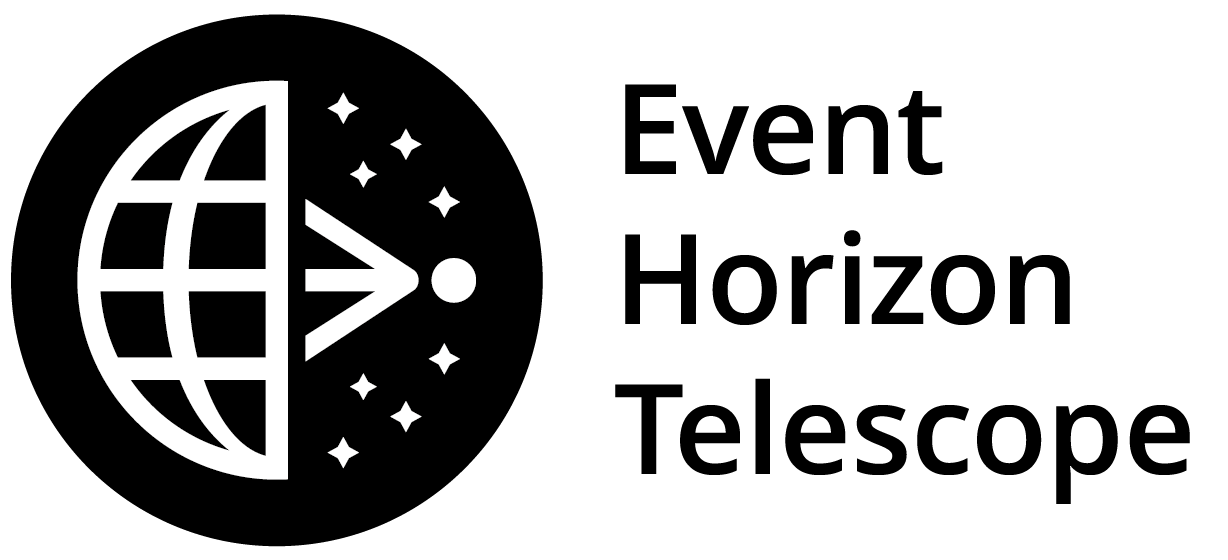}}
    \label{fig:my_label}
\end{minipage}
\begin{minipage}[t]{0.49\textwidth}
\begin{flushright} \large
\textbf{Event Horizon Telescope \\ Memo Series} 
\end{flushright}
\end{minipage}
\end{figure}

\begin{center}
EHT Memo \memonr{}
\end{center}

{\let\newpage\relax\maketitle}  
\thispagestyle{empty}		

\begin{flushleft}
{\small
\affiliations{}
}
\end{flushleft}

\begin{center}
\textbf{Abstract}
\end{center}

\justifying

This document presents a novel method for the intra-field calibration and imaging of the radio source \sgra, observed with the Atacama Large Millimeter/submillimeter Array (ALMA).  \sgra  is a complex source comprising two components: the compact core (which exhibits high variability) and the extended minispiral (which is relatively stable over short timescales). 
The novel approach consists in a self-calibration method that employs the extended structure of the source (the minispiral) to calibrate the flux variability of the compact core. The algorithm\footnote{You can download the full pipeline in github, using the link: \url{https://github.com/ealruiz/calminispiral}.} involves several steps: (1) an initial CLEAN image is generated for the entire source; (2) the core is subtracted, leaving only the minispiral; (3) a two-component visibility model is constructed, comprising the minispiral and the core; (4) the model is fitted to the data, retrieving flux density parameters for each integration time; and (5) the data are scaled and calibrated, resulting in nearly constant brightness for the minispiral and variable flux for the core.
The implementation of this algorithm 
through a script in the CASA package is described, detailing the configuration parameters and the steps involved. The success of the method is demonstrated through light curves of \sgra  observed on day 21 April 2018 in Band 6, as part of the 2018 EHT campaign. The light curves have been produced for Stokes I (total intensity), linear polarized intensity, EVPA (electric vector position angle), and Stokes V (circular polarization), providing valuable insights into the variability of this radio source.

\vspace{\baselineskip}

\newpage

\tableofcontents

\section{Introduction}
\label{sec:theory}

ALMA data acquired during VLBI observations are calibrated in a process known as “Level 2 Quality Assurance” (QA2).  
This process is based on self-calibration assuming a point source with a constant flux density.
While this approach is well justified (and effective) for most VLBI targets, it is not optimal for Sgr A*. 
In fact, \sgra  can be understood as the sum of two components:
\begin{itemize}
\item An extended component (up to parsec scales), known as the  {\bf minispiral}. Given its large extension, it is safe  to assume that this component is not variable on the timescales comparable to the duration of the observations (i.e., a few hours).
\item A compact, bright component, which presents high variability: {\bf \sgra itself}. The flux density of the \sgra core is higher than the total flux density  of the  minispiral.
\end{itemize}

The constant flux density assumption of the QA2 calibration, 
results in a core with a constant brightness while transfers all the variability to the minispiral \footnote{If we could make a movie of the Fourier Transform (FT) of the QA2  calibrated visibilities, we would get a core with almost constant flux, while the brightness of all the minispiral would be pulsating (the flux decreases and increases at short timescales), as opposed to the true behaviour of the source.}.

In order to derive the light curves of \sgra (i.e. the compact core), we designed and implemented a novel algorithm to extend and improve  
the QA2 gain calibration for variable sources.

\section{Algorithm description}

Our algorithm consists in an {\it intra-field} calibration and imaging of ALMA observations of \sgra. The algorithm is already implemented in a script in CASA (see sec. \ref{sec:script}) and consists of 4 main steps:
\begin{itemize}
\item {\bf Step 1:} CLEAN image

First, we take all the visibilities acquired in the same track and calibrated in the QA2 process 
and we generate an image of  the source (i.e. the core and the minispiral).  This image, which we indicate with \texttt{IM0}, typically presents  artefacts (e.g. negative artefacts that mimic partial sidelobes of the PSF around \sgra), but  serves as an initial approximation to the minispiral.
\end{itemize}

\begin{itemize}
\item {\bf Step 2:} Separate models

We subtract the core of \sgra  from the image space (\texttt{IM0}) to get an image of only the minispiral (\texttt{IM1}), by selecting the pixels of the core (strongest CLEAN component) and choosing one of the two following approaches:
\begin{itemize}
\item set the flux of the pixels to 0,
\item replace the flux of the pixels to the extended brightness of the minispiral (i.e. the average of the neighbouring pixels) to get a continuation of the minispiral where the core was located.
\end{itemize}
Both approaches give compatible results after the first iteration of the algorithm. 

\end{itemize}

\begin{itemize}
\item {\bf Step 3:} Visibility (two-component) model fitting

Next, we construct a Visibility model from the \texttt{IM1} CLEAN components, as the FT of the minispiral ($ \texttt{MOD}=FT(\texttt{IM1})$) scaled by a time-varying factor $S_1$ (due to the variability of the minispiral's total flux density\footnote{Note that the minispiral is introducing time-dependent changes on the visibilities, which come from the source structure and are absorbed by the FT; therefore $S_1$ only absorbs the time variation due to the variability of the minispiral's total flux density.}) and a constant factor $S_2$ (i.e. the FT of a point source at the center of the image):
\begin{equation}
V^{mod} = S_1 \cdot \texttt{MOD} + S_2.
\end{equation}
\end{itemize}
This visibility model describes the entire structure of \sgra, with a fixed minispiral (\texttt{IM1}) and the core (i.e. a centered point source). 
For each integration time,  we fit  the observed visibilities to our visibility model, that is, we minimize the $\chi^2$ (error function)
\begin{equation}
\chi^2(t_{int})=\sum_{i,t=t_{int}} \omega_i \cdot \abs{V_i^{mod}(t) - V_i^{obs}(t)}^2 = \sum_{i,t=t_{int}} \omega_i \cdot \abs{ S_1 \cdot \texttt{MOD}(t) + S_2 - V_i^{obs}(t)}^2,
\label{eq:chi2}
\end{equation}
where $\omega_i$ are the statistical weights of each visibility. The fitting parameters (different for each integration time $t_{int}$) are $S_2$ (the flux of the core) and $S_1$ (a scale factor for the minispiral, which accounts for the artificial variability introduced by the QA2).

Our approach is the standard least-squared minimization of the $\chi^2$ function:
\begin{align}
	\frac{\partial \chi^2}{\partial S_m} & = \frac{\partial }{\partial S_m} \left[\sum_{i} \omega_i \left(S_1 \cdot \texttt{MOD}_i + S_2 - V_{i}^{obs}\right) \left(S_1 \cdot (\texttt{MOD}_i)^{\ast} + S_2 - (V_{i}^{obs})^{\ast}\right)\right] = 0; \\
        \frac{\partial \chi^2}{\partial S_1} & = 2 \sum_{i} \omega_i \left[\abs{\texttt{MOD}_i}^2 S_1 + Re\{V_i^{mod}\} S_2 - \left(Re\{\texttt{MOD}_i\}Re\{V_i^{dat}\} + Im\{\texttt{MOD}_i\}Im\{V_i^{dat}\} \right) \right]= 0; \\
        \frac{\partial \chi^2}{\partial S_2} & = 2 \sum_{i} \omega_i \left[Re\{\texttt{MOD}_i\} S_1 + S_2 - Re\{V_i^{dat}\} \right]= 0.
    \label{eq:chi2_grad}
\end{align}
Since the $\chi^2$ is linear for the fitting parameters $S_k$, we can easily retrieve the algebraic expression of the Hessian $H$ and the Residual vector $\vec{b}$,  corresponding to the matrix formulation of the least-squared minimization problem, with Parameter vector $\vec{S}$:
\begin{equation}
H\vec{S} = \vec{b}.
\label{eq:matrix_model_fitting}
\end{equation}
Using this formalism, from eqs. \ref{eq:chi2_grad}, we retrieve:
\begin{align}
H &= 
\begin{pmatrix}
2 \sum_{i} \omega_i \abs{V_i^{mod}}^2 & 2 \sum_{i} \omega_i Re\{V_i^{mod}\} \\
2 \sum_{i} \omega_i Re\{V_i^{mod}\}     & 2 \sum_{i} \omega_i
\end{pmatrix}, \\
\vec{b} &= 
\begin{pmatrix}
2 \sum_{i} \omega_i Re\{V_i^{mod}\}Re\{V_i^{dat}\} + Im\{V_i^{mod}\}Im\{V_i^{dat}\} \\
2 \sum_{i} \omega_i Re\{V_i^{dat}\}
\end{pmatrix}.
\end{align}

{\bf What about polarization?} As the minispiral polarization is negligible (due to \sgra dominating both in total intensity and polarization brightness), the model for Stokes $Q,U,V$ is a centered point source (i.e. $V^{mod}=S$, being $S$ the flux of the core Stokes parameter). Under this assumption, it can be shown (with the same least-square minimization approach used for Stokes $I$) that $S$ is the average of the real part of the visibilities.

\begin{itemize}
\item {\bf Step 4:} Scale data \& Calibrate light curves

We retrieve, from the model-fitting (step 3), $S_1$  and $S_2$. Since we fit our model to the data from the QA2, we retrieve a $S_2$ with low variability, while $S_1$ changes continuously. 
In order to transfer the variability from $S_1$ to $S_2$, we can write the QA2 visibilities as a function of time $t$  as:
\begin{equation}
V_{QA2}(t) = K(t) \cdot V_{cal}(t) \rightarrow V_{cal}(t) = \frac{\bar{S_1}}{S_1(t)} \cdot V_{QA2}(t),
\label{eq:lcurve_scale}
\end{equation}
where $ V_{cal}(t)$ are the properly calibrated visibilities, and $\bar{S_1}$ is the median of all the flux densities for all the frames of the minispiral (i.e. the median of all $S_1(t)$ across all the days of the campaign\footnote{This additional gain to the QA2 visibilities ensures consistency when scaling the light curve fluxes of the different days, by imposing the assumption of a constant flux of the minispiral along the campaign.}). This means that, when in our fitting procedure the brightness of the minispiral increases with respect to \texttt{IM1} (the fixed minispiral), we scale the minispiral to decrease $V_{cal}(t)$, and if the brightness decreases, we scale the minispiral increasing $V_{cal}(t)$. As a result, we are correcting the amplitude of all the visibilities for each integration time, so that the minispiral brightness remains (almost) constant, while the core showcases the expected flux variability. 

Moreover, these scale factors (or gains $\bar{S_1}/S_1(t)$) are the same for all the Stokes parameters, since they take into account the atmospheric opacity and the antenna gains, while the X-Y ratio is assumed to be constant. Therefore,  the gains are polarization independent, and can be applied equally to all the visibilities (\texttt{XX, XY, YX, YY}). As a result, we obtain both the Stokes $I$ and the polarized light curves.

\end{itemize}

{\bf What does it means to iterate?}

Once we have calibrated the visibilities, we retrieve two flux components: \sgra and the minispiral, as $V_{cal}=V_{minispiral}+V_{SGRA\star}$, for each integration time. Using these flux estimates, we can update the visibility model by subtracting the estimated flux of \sgra. Since \sgra is a point source, its flux corresponds to a constant in Fourier space (i.e., the same value added to all visibilities). Therefore, by subtracting this constant from the visibilities for each integration time, we isolate the Fourier Transform of the minispiral. These minispiral visibilities are then rescaled to the average flux density, $\bar{S_1}$.

With these residual visibilities, which now correspond solely to the signal of the minispiral, as the \sgra flux has been removed, we can proceed with the imaging to retrieve an improved image of the minispiral without \sgra contamination. This updated image can be used as a refined minispiral model (\texttt{IM1}), so we can repeat the iterative steps to ensure the convergence of this method. In this new iteration, when fitting the corrected visibilities (i.e., applying eq. \ref{eq:chi2} with $V_{obs}(t) = V_{cal}(t)$), we now retrieve a constant $S_1$ while $S_2$ captures all the variability, indicating proper calibration.

\section{Description of the intra-field \sgra ALMA calibration script}
\label{sec:script}

The code performs a series of steps to calibrate and analyse radio interferometry data, focusing on the compact component (\sgra) and the extended component (the minispiral) of the observed source\footnote{We use the \texttt{CASA} (Common Astronomy Software Applications) package, version 4.7.0 (\url{https://casa.nrao.edu/download/distro/casa/release/el7/casa-release-4.7.0-el7.tar.gz}).}. For the first iteration of the intra-field ALMA calibration, the steps 1-4 must be run;  step 5 is optional, and can be run to re-CLEAN the minispiral and self-calibrate.

\subsection{Configuration}

The code begins by setting up various configuration parameters needed for each step. Here we summarize the relevant parameters for each step:

\begin{itemize}

\item {\bf Step 1:}
\begin{itemize}
\item The name of the files need to be set, for each track, as "\textit{$\{\texttt{DATNAM}\}$\_$\{\texttt{track}\}$.ms}", where:
\begin{itemize}
\item \texttt{DATNAM}: Common name of all measurement set files.
\item \texttt{TRACKS}: List of tracks where \sgra was observed.
\end{itemize}
\item \texttt{IMSIZE}: Image size (number of pixels) for the CLEAN process.\item \texttt{Ns}: Nyquist sampling parameter.
\item \texttt{Bmax}: The greatest projected baseline length, representing the highest resolution.
\item \texttt{Cell}: Cell size for imaging, calculated based on Nyquist sampling and \texttt{Bmax}.
\item \texttt{CELL}: Cell size, formatted as a string.
\end{itemize}

\item {\bf Step 2:}
\begin{itemize}
\item \texttt{REMOVE\_CENTER}: Boolean flag indicating whether to remove \sgra  from the extended model (recommended to be \texttt{True} to study the minispiral of \sgra).
\item \texttt{REMOVE\_ALL\_CENTER\_BEAM}: Boolean flag indicating whether to set the central pixel to zero or the average in-beam extended brightness.
\end{itemize}

\item {\bf Step 3:}
\begin{itemize}
\item \texttt{MINBAS}: Minimum baseline length in meters for flagging short baselines (not used in our approach).
\item \texttt{RELOAD\_MINISPIRAL\_MODEL}: Boolean flag indicating whether to reload the model of the extended component (recommended to be \texttt{True}).
\item \texttt{USE\_SELFCAL\_DATA}: Boolean flag indicating whether to use self-calibrated data; only \texttt{True} if repeating this step, after step 5.
\end{itemize}

\item {\bf Step 4:}
\begin{itemize}
\item \texttt{SGRA\_MIN, SGRA\_MAX}: Minimum and maximum allowed \sgra  flux densities for flagging bad integrations.
\item \texttt{SGRA\_MAX\_PolI}: Maximum allowed \sgra  polarized intensity, for flagging bad integrations.
\item \texttt{EXPORT\_MINISPIRAL}: Boolean flag indicating whether to export the calibrated minispiral visibilities, scaled to the flux-density median; needed if step 5 will be run.
\item \texttt{CORRECT\_SPWS2\_B6}: Boolean flag indicating whether to correct for the lower \sgra flux density at ALMA B6 spectral window 2, caused by the absorption line at 227 GHz. Both the absorption line and the details on the correction model are described in Ap. \ref{sec:absorption}.
\end{itemize}

\item {\bf Other variables:}
\begin{itemize}
\item \texttt{mysteps}: List of steps to be executed; required: steps 0-4; optional: step 5 and repeat steps 3-4.
\item \texttt{thesteps}: Dictionary mapping step numbers to their corresponding descriptions.
\end{itemize}

\end{itemize}

\subsection{Step 0}
This step uses the \texttt{split} task to store, from the data set obtained after the QA2 calibration, only the \sgra  visibilities on a new \texttt{'*.ms'} data set\footnote{Note that, for optimum speed, there should be only one channel per spw on the \texttt{'*.ms}}, and adds a model column to the \texttt{'*.ms'} (using the task \texttt{clearcal}).

\subsection{Step 1}
\begin{figure}[h!]
\centering
    \includegraphics[width=12.5cm]{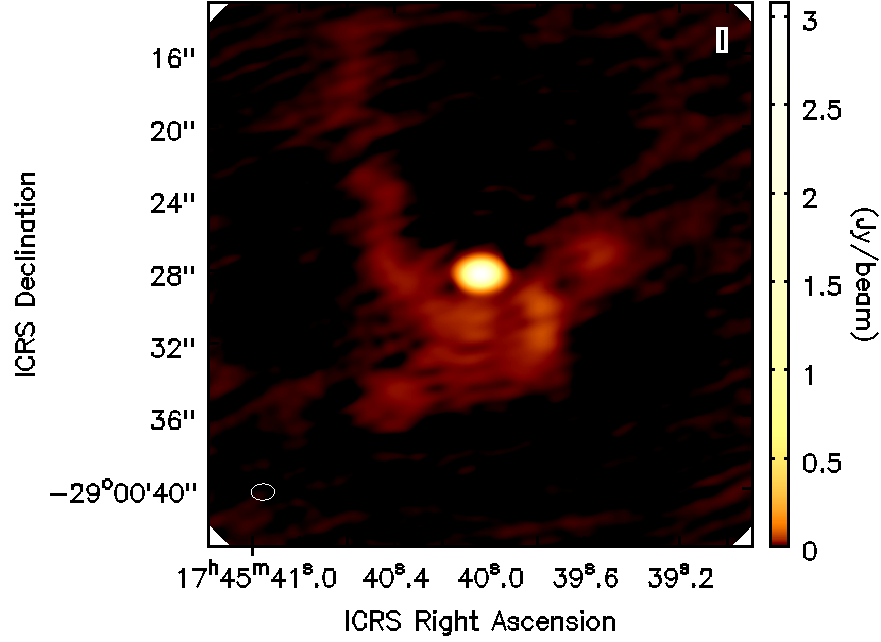}
     \caption{Stokes I CLEAN image of \sgra  + the minispiral from the visibilities produced after the QA2 of the ALMA observations, track E18E22, for the spw 0.}
     \label{fig:CLEAN_SGRA_image}
\end{figure}

This step performs the first CLEANing (\texttt{tclean} task) of both the compact and extended components (\sgra  and the minispiral), iterating over spectral windows (the CLEAN mask is improved incrementally), and generates a \texttt{'*.model'} image (the CLEAN components), with both \sgra  and the minispiral (i.e. the image \texttt{IM0}, see Fig. \ref{fig:CLEAN_SGRA_image}).

\subsection{Step 2}
\begin{figure}[h!]
\centering
    \includegraphics[width=12.5cm]{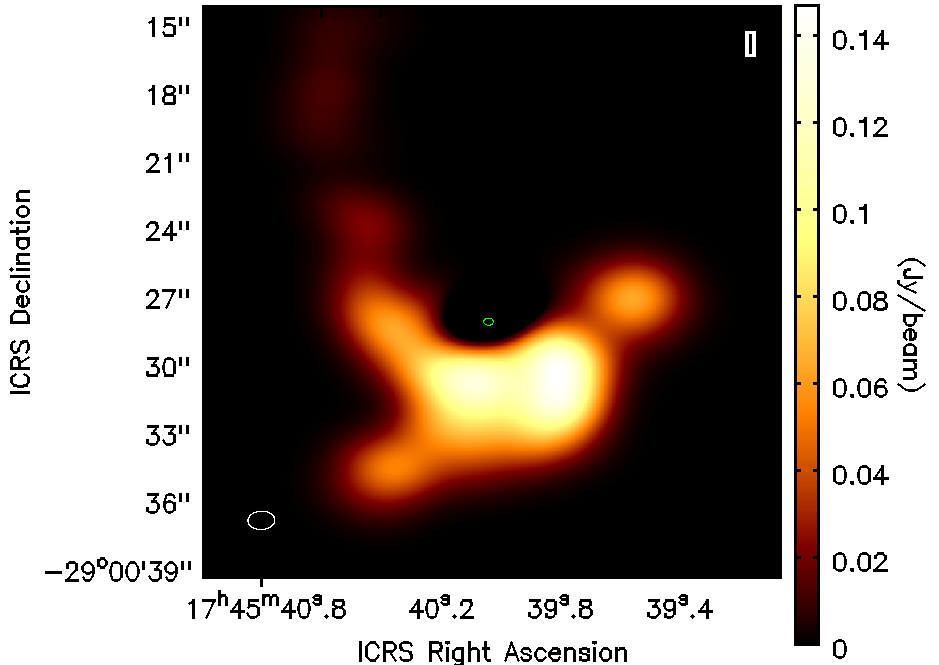}
     \caption{Stokes I CLEAN image of the minispiral after removing, for each spw, the compact component from the \texttt{'*.model'} produced by step 1, and averaging all spw. The green contour (at 0.9 times the \sgra core Stokes I peak) marks the position of the removed core.}
     \label{fig:CLEAN_noPeak_image}
\end{figure}

This step separates, for each spectral window (spw), a source MODEL (CASA \texttt{'*.model'}) into two components: a centered point source and an extended source:
\begin{itemize}
\item reloads the image metadata from the CLEAN \texttt{'*.image'} (beam size, position angle and pixel size) and computes the beam area (in pixel units),
\item reads the CLEAN components (from the \texttt{`*.model'} image), gets the coordinates of the   CLEAN component with strongest flux density, and removes the centered point source:
\begin{itemize}
\item either by removing the whole flux density of the peak pixel completely, or
\item by making it equal to the average extended surrounding brightness, estimated from the average clean flux-density (local brightness) surrounding the inner clean beam in the image center, after subtracting the flux-density of the point source contribution within the mask).
\end{itemize}
\item convolves the resulting model (i.e., minispiral without the central compact component) with the CLEAN beam
\item finally, this step computes an average CLEAN model over all four spws, to get the average extended structure (\texttt{IM1}, see fig. \ref{fig:CLEAN_noPeak_image}), exports \texttt{IM1} into a new CASA \texttt{'*\_noPeak.image'}, and writes its CLEAN components (\texttt{RA}, \texttt{dec} and flux density) in an ascii \texttt{'*\_CCs.dat'} file.
\end{itemize}

\subsection{Step 3}

This step computes the FT of all the CLEAN components of the extended source, loaded from the ascii file, and estimates the flux of the source, given as the sum of an extended and a compact component (parameters $S_1$ and $S_2$, respectively), at each integration time and spectral window. The fitting is done algebraically (see eq.\ref{eq:matrix_model_fitting}) by inverting the (2x2) Hessian matrix, since the model is linear for our parameters. This step stores, for each spw, all these results into a numpy named array, which is pickled together with the frequency of the corresponding spw, into an external \texttt{'*.fit'} file.
\begin{itemize}
\item Description of the \texttt{FIT\_ARRAY} stored into the \texttt{'*.fit'} file:
  \begin{itemize}
   \item \texttt{FIT\_ARRAY[k]} access the named columns for each spw
   \item named columns: \texttt{'JDTime', 'I Extended', 'I Compact', 'I LongBas', 'Q', 'U', 'V', 'Error Ext.', 'Error Comp.', 'Covariance', 'Error Q', 'Error U', 'Error V'}. 
   \end{itemize}
   An additional \texttt{'GOODS'} column is included, comprising \texttt{True} boolean values to denote times with outliers identified during light curve inspection, to be marked as \texttt{False}.
\item This step also updates the \texttt{`Corrected data'} column of the \texttt{'*.ms'}, so that it only contains the extended component data (i.e. the minispiral), after subtracting the compact flux from the data and imposing that the extended component has a constant flux, normalized (scaled) to 1Jy.
\end{itemize}

\subsection{Step 4}

\begin{figure}[h!]
\centering
    \includegraphics[width=16.5cm]{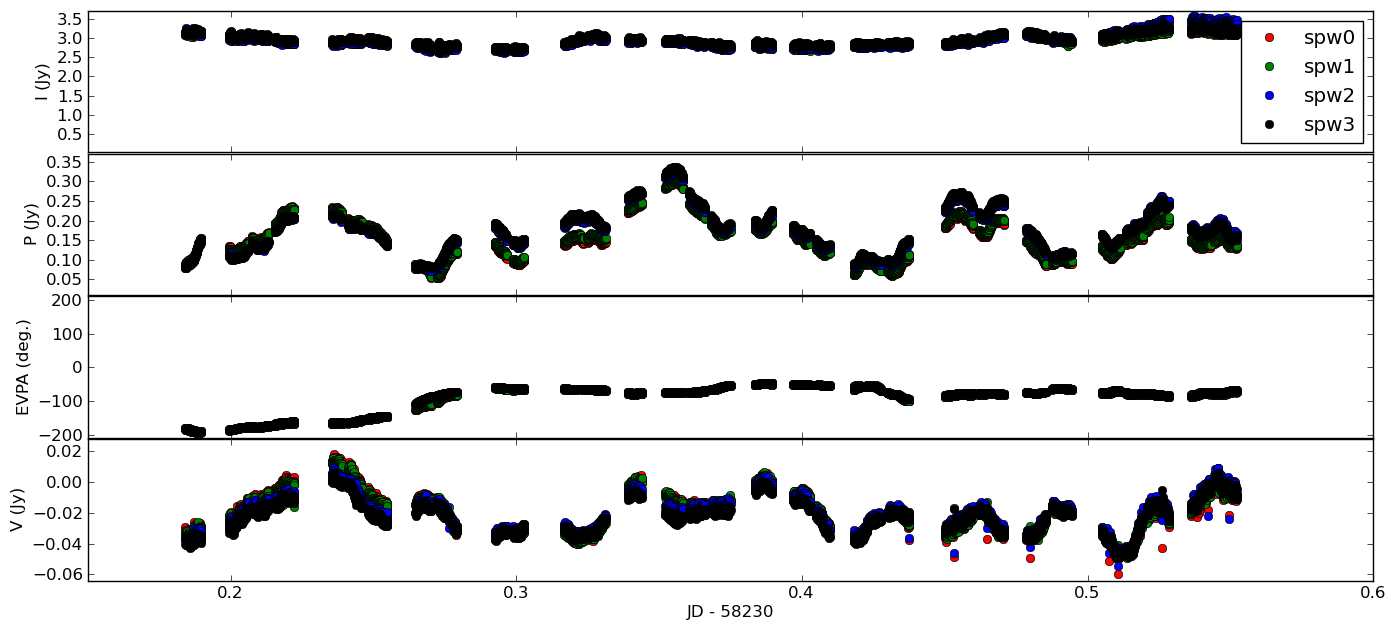}
     \caption{Light curves of \sgra after intra-field calibration of the April 22, 2018, ALMA observations. From top to bottom, the figure show Stokes $I$, the linear polarized intensity $P$, the EVPA and Stokes $V$.}
     \label{fig:lightcurvesSGRA}
\end{figure}

The light curves of \sgra  are calibrated (see fig. \ref{fig:lightcurvesSGRA}) by scaling the corrected data to the median flux density of the minispiral source, by computing the gains (i.e. $\bar{S_1}/S_1$ at each integration time, for each spw). This step also saves, for each spw, the calibrated light curves in an ascii \texttt{'Light\_Curve\_SPWi\_*.dat'} file, with the following info.: \texttt{MJD, I(Jy), Ierr(Jy), P(Jy), Perr(Jy), EVPA(deg), EVPAerr(deg), V(Jy), Verr(Jy)}.
\begin{itemize}
\item If \texttt{EXPORT\_MINISPIRAL} is set to \texttt{True}, splits the minispiral-only calibrated data (i.e. the \texttt{`Corrected data'} column) to a \texttt{'*.ms\_ExtendedData'} measurement set, and scales the minispiral data to the flux-density median.
\end{itemize}

Note that this step scales the light curves by employing the intra-day average of the minispiral flux. However, for ensuring consistency throughout the entire campaign, it is advisable to scale the flux of the light curves using the median value derived from the minispiral flux observed across all days. Additionally, it is recommended to manually identify and remove any outliers from the light curves, and subsequently mark the the \texttt{'GOODS'} column of the \texttt{'*.fit'} file to \texttt{False}.

\subsection{Step 5 (OPTIONAL Iteration)}

This optional step Re-CLEANs the minispiral, self-calibrates the visibilities, and writes the average minispiral image to an ascii \texttt{'*\_CCs.dat'} file. To run this step, \texttt{EXPORT\_MINISPIRAL} should have been set to \texttt{True} (at the step 4 config.). After running step 5, set \texttt{USE\_SELFCAL\_DATA} to \texttt{True} (at the step 3 config.), \texttt{EXPORT\_MINISPIRAL} to \texttt{False} (at the step 4 config.), and run steps 3-4.

\section{Products of the minispiral calibration for use in VLBI}

Step 3 of our algorithm provides a \texttt{'*.fit'} file with the time-varying factor $S_1$, located in the \texttt{'I Extended'} column, reflecting the variability of the minispiral's total flux density. This information must be transferred back to \sgra, following the subsequent Step 4 of our calibration scheme. Additionally, the constant factor $S_2$, corresponding to the Stokes parameters stored in the \texttt{'I Compact', 'Q', 'U', 'V'} columns without the variability, is also included in this file.

Moreover, after calibrating the light curves for both total intensity and polarization, we can identify instances of clear outliers in the data (which we indicate as a \texttt{False} boolean in the \texttt{'GOODS'} column of the \texttt{'*.fit'} file). Subsequently, we convert the timestamps of these outliers into a format suitable for CASA tasks, allowing the flagging of outliers in the visibility space. The timestamp format used is YYYY/MM/DD/HH:mm:ss, within a window of $\pm 1s$, and these flagged outliers are stored in a \texttt{'*.dat'} file for further analysis.

\section{Conclusions.}
In conclusion, we introduce an approach for the intra-field ALMA calibration and imaging of the radio source \sgra, located at the center of the Milky Way galaxy. By leveraging the extended structure of the source (the minispiral) to calibrate the flux variability of the compact core, this method addresses the challenge of amplitude gain calibration, where the existing QA2 assumes a constant flux density for \sgra.

The algorithm involves a series of steps, including CLEAN imaging, core subtraction, two-component visibility modeling, data fitting, scaling and calibration. This self-calibration method allows for the accurate characterization of light curves for both \sgra  and the minispiral, providing valuable insights into the behavior of this enigmatic radio source.

The implementation of the approach in the CASA package is described, with detailed configuration parameters and step-by-step instructions. The resulting light curves of \sgra, including Stokes I, linear polarized intensity, EVPA, and Stokes V, demonstrate the effectiveness of the proposed method in capturing the variability of the core while maintaining a nearly constant brightness for the minispiral.


\begin{appendix}

\section{Absorption line at 227 GHz}
\label{sec:absorption}

\begin{figure}[h!]
    \centering
    \includegraphics[width=11cm]{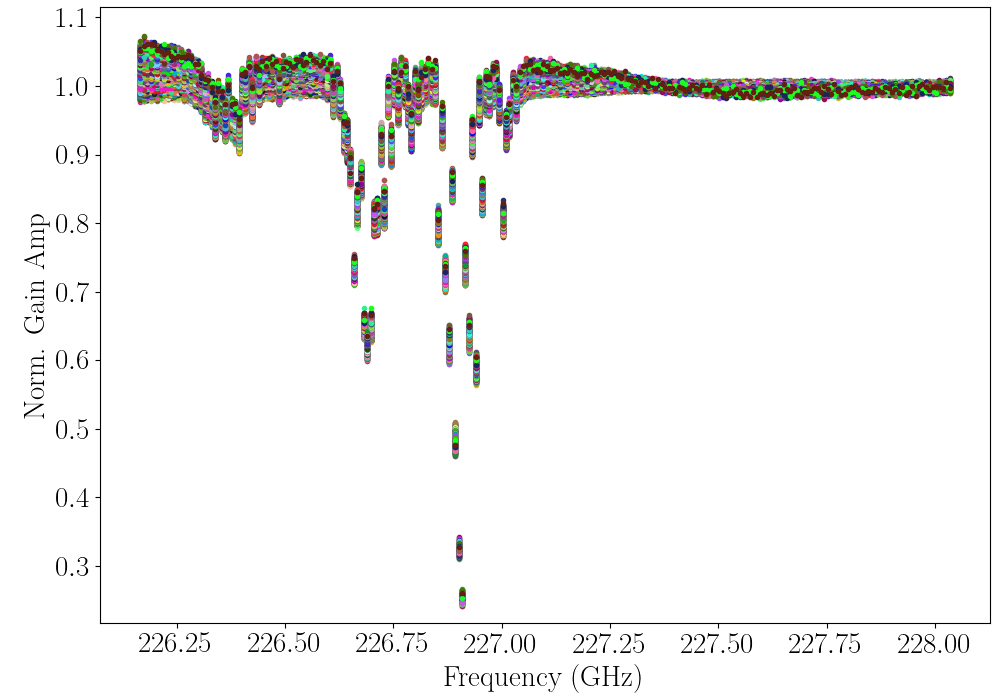}
    \caption{Bandpass of \sgra for the April 22 observations, indicating the channels impacted by absorption at spw 2. The bandpass remains virtually unchanged across all days.}
    \label{fig:absorption_chan_spw2}
\end{figure}

As depicted in Fig. \ref{fig:absorption_chan_spw2}, the spectrum of \sgra in the spw 2 (centered around 227 GHz) on April 22 exhibits absorption at channels 17-33 and 56-112, consistent across all days of the campaign. This foreground absorption toward \sgra at 226.91 GHz mirrors that reported in the ALMA 2017 observations \citep[see][]{Goddi2021}, albeit more abrupt, and leads to a flux decrease in this spw. Consequently, the frequency channels affected by absorption must be flagged during QA2 calibration to ensure consistency in the light curves of \sgra across all spws.

\end{appendix}

\newpage

\pagebreak

\pagebreak


\begin{thebibliography}{9}

\bibitem{Goddi2021}
Goddi, Ciriaco, et al. 
``Polarimetric Properties of Event Horizon Telescope Targets from ALMA.'' 
\textit{The Astrophysical Journal Letters}, vol. 910, no. 1, 2021, p. L14. 
DOI: \href{https://doi.org/10.3847/2041-8213/abee6a}{10.3847/2041-8213/abee6a}.

\end{thebibliography}
\end{document}